\documentclass[aps,prc,groupedaddress,showpacs,showkeys,
nofootinbib,fleqn]{revtex4}
\usepackage{epsfig,amsmath,amssymb,graphicx,tabularx,dcolumn,bm,longtable}
\usepackage{verbatim}  % cor comments

\begin{document}

\title{Gamow-Hartree-Fock-Bogoliubov Method: Representation of quasiparticles
with Berggren sets of wave functions}

\author{N.~Michel}
\affiliation{CEA, Centre de Saclay, IRFU/Service de Physique Nucl{\'e}aire, F-91191 Gif-sur-Yvette, France}
\author{K.~Matsuyanagi}
\affiliation{Theoretical Nuclear Physics Laboratory, RIKEN Nishina Center,  
Wako 351-0198, Japan} 
\author{M.~Stoitsov}
\affiliation{Department of Physics and Astronomy, University of Tennessee, Knoxville, Tennessee 37996, USA}
\affiliation{Physics Division,  Oak Ridge National Laboratory, P.O.~Box 2008, Oak Ridge,Tennessee 37831, USA}
%\affiliation{Joint Institute for Heavy-Ion Research, Oak Ridge, Tennessee
%37831, USA}
\affiliation{Institute of Nuclear Research and Nuclear Energy, Bulgarian Academy of Sciences, Sofia-1784, Bulgaria}

\date{\today}

\begin{abstract}
Single-particle resonant states, also called Gamow states, 
as well as bound and scattering states of complex energy
form a complete set, the Berggren completeness relation. 
It is the building block of the recently introduced Gamow Shell Model, 
where weakly bound and resonant nuclear wave functions are expanded 
with a many-body basis of Slater Determinants generated by this set of 
single-particle states. 
However, Gamow states have never been studied in the context
of Hartree-Fock-Bogoliubov theory, except in the Bardeen-Cooper-Schriefer 
(BCS) approximation, where both the upper and lower components 
of a quasiparticle wave function are assumed to possess the same radial 
dependence with that of a Gamow state associated with the Hartree-Fock 
potential. Hence, an extension of the notion of Gamow state has to be effected 
in the domain of quasiparticles.
It is shown theoretically and numerically that bound, resonant and 
scattering quasiparticles are well defined and form a complete set, 
by which bound Hartree-Fock-Bogoliubov ground states can be constructed.
It is also shown that the Gamow-Hartree-Fock single-particle basis  
can be used to solve the Gamow-Hartree-Fock-Bogoliubov problem. 
As an illustration, the proposed method is applied to neutron-rich 
Nickel isotopes close to the neutron drip-line.
\end{abstract}

\maketitle

\section{Introduction}
\label{Intro}
One of current challenges of nuclear theory is the quantitative description of 
nuclei situated near and beyond drip-lines.
Powerful facilities are being built in several countries in order to 
create these very short-lived states.
%such as HRIBF in RIKEN (Japan), GSI upgrade with FAIR (Germany), 
%ISAC I-II in TRIUMF (Canada) and SPIRAL I-II in GANIL (France).
%Ambitious projects are also studied with EURISOL in Europe, which gathers many %laboratories and universities, and RIATG in the USA.
For a long time, microscopic theories of nuclear structure have been developed 
for describing ground states of nuclei close to the valley of stability.
For describing stable nuclei which are well localized, the harmonic oscillator 
(HO) bases are useful for both shell model \cite{SM_review} 
and Hartree-Fock Bogoliubov (HFB) calculations 
\cite{Gog75, Gir83, Egi80, Egi95, Dob04}; the HO bases converge quickly therein.
However, it possesses poor convergence properties for weakly bound nuclei 
bearing large spatial extensions, which lie very close to neutron drip lines. 
%Situation is even worse for particle-unstable ground states 
%beyond the particle drip line, such as proton emitters, 
%as they do not belong to the Hilbert space,
%and are thus theoretically unattainable with the HO basis. 
%Use of complex-scaled Hamiltonians 
%\cite{Kruppa_Kato} can render resonant states integrable,
%so that their complex-scaled representations belong to the Hilbert space.
%Nevertheless, the price to be paied in this method is that one must impose 
%exact asymptotic of many-body states
%with Jacobi coordinates \cite{Csoto}, which is possible only for very 
%lightest nuclei or in a few-cluster approach \cite{Csoto,Descouvemont}.

A promising approach to this problem has been proposed 
in Refs.~\cite{PRL_michel,PRC1_michel,PRC2_michel} 
within a shell model framework; namely, the Gamow Shell Model (GSM).
The fundamental idea is to replace the HO basis by the Berggren basis 
consisting of bound states, resonance states and continuum scattering 
states of complex energy, generated by a single-particle potential. 
It has been shown numerically that this basis has the ability to expand 
both halo nuclei and many-body resonant states precisely. 
The latter indeed belongs to a rigged Hilbert space 
\cite{R_de_la_Madrid1,R_de_la_Madrid2}, 
which is an extension of the notion of Hilbert space to non-square integrable 
wave functions.
However, the dimension of the Berggren Slater Determinants 
represented by the GSM basis increase 
very quickly with increasing number of valence particles; 
it increases much faster than in standard shell model
due to the presence of occupied scattering states. Hence, the GSM is a tool 
mainly dedicated to the study of light nuclei. For medium and heavy nuclei,
a method of choice is the HFB, which can be followed by quasiparticle random 
phase approximation (QRPA). As pairing correlations 
are absorbed in the HFB ground state, one-body nature of the 
HFB framework enables fast evaluations of ground states 
of medium and heavy nuclei, and it is in fact the only method 
suitable for systematic calculations; 
see Ref.~\cite{sto04} for an evaluation of even-even nuclei in  
the whole nuclear chart with the HFB formalism.
In order to properly treat drip-line nuclei within the HFB framework, 
the real-space coordinate-mesh method has been applied 
using box boundary conditions 
\cite{Bul80, Dob84}. 
Extension of this approach to deformed nuclei is difficult
and has been carried out only recently \cite{Ter03, Obe03}. 
As an alternative more convenient approach, 
one can adopt basis expansion methods, where direct integration procedure is 
replaced by matrix diagonalization. 
A first amelioration of the HO basis had been proposed with the transformed 
harmonic oscillator (THO) basis \cite{sto04,hfbtho}. 
Applying unitary transformations to the HO basis, one obtains the THO basis, 
in which Gaussian fall-off of the HO wave functions is replaced by physical 
exponential decrease of the THO basis wave functions. 
However, the THO basis always dictates exponential decrease in 
expanding quasiparticle wave functions,
for both upper and lower components, 
even when they are part of scattering states, 
so that unsatisfactory basis dependence remains. 
In order to solve this problem, a new basis has been introduced very recently, 
which consists in using bound and continuum basis states 
generated by the analytic 
P{\"o}schl-Teller-Ginocchio (PTG) potential~\cite{PTG_pot}. 
The PTG basis introduced in this paper~\cite{HFB_PTG_michel} 
possesses a peculiarity to bear no narrow resonance states; 
those are replaced by bound PTG states.
Thus, PTG continuum set of basis states can be discretized very effectively 
with Gauss-Legendre quadrature, as they contain no resonant structure.
It has been shown that they can provide a good description of spatially 
extended nuclear ground states of both spherical and axially deformed 
nuclei~\cite{HFB_PTG_michel}.
On the other hand, the PTG basis formed by bound and real scattering states 
is not a Berggren complete set of states, 
so that it would be more convenient to use a Berggren quasiparticle basis set, 
when we are interested in describing particle-decaying excited states. 
Up to now, however, resonant quasiparticle states have been studied in the 
context of Berggren completeness relation 
only within the BCS approximation \cite{Betan,Dussel}.
The last approach is indeed not satisfactory due to the well-known gas 
problem arising from the occupation of the continuum: In fact, densities are 
not localized in the BCS approach, because the lower components of scattering 
quasiparticle states are of scattering type as well.
Contrary to what is stated in Ref.~\cite{Betan}, it cannot be regularized 
using complex scaling because it does not have pure outgoing asymptotic. 
Use of continuum level density in Ref.~\cite{Dussel} is also problematic, 
even though it suppresses the gas problem. Indeed, it is not part of
continuum HFB theory~\cite{HFB_PTG_michel}, so that its introduction in
HFB equations strongly modifies quasiparticle coupling to the continuum.
In particular, it supresses a large part of non-resonant continuum, and thus
important physical properties of drip-line nuclei as well. Hence, with this approach, 
weakly bound systems cannot be studied properly. Only a full application of 
the HFB framework can unambiguously solve this problem, where densities are 
localized by construction for bound HFB ground states.

The major purpose of this paper is to develop a new method of solving the 
continuum HFB equations utilizing the Berggren basis,
called Gamow-HFB method, by which 
bound, resonant and continuum quasiparticle states are provided.  
It allows expansion of QRPA excited states having escaping widths 
in terms of the Berggren quasiparticle basis associated with the bound 
HFB ground state. 
This is very important because, in weakly bound unstable nuclei, 
low-lying collective excited states may acquire particle-decay widths.

This paper is organized as follows.
Firstly, the standard HFB formalism is briefly summarized. 
As we use the Skyrme interactions \cite{Vautherin_Skyrme}, 
it is effected in the context of density functional theory (DFT). 
Secondly, we define quasiparticle S-matrix poles and scattering states 
of complex energy; 
these are direct extensions of their single-particle counterparts.
We then present the quasiparticle Berggren completeness relation 
generated by those states. Numerical methods to calculate Gamow 
and complex scattering quasiparticle states are described;
they differ significantly from the scattering quasiparticle states 
discretized by box boundary conditions. 
We also present another method of solving the continuum HFB equations 
in which the HFB quasiparticle wave functions are expanded in terms of 
the Gamow-Hartree-Fock (GHF) basis; this approach may be regarded as  
an extension of the standard two-basis method \cite{Gal94, Ter97a, Yam01} 
to complex energy plane. 
Feasibility of the proposed methods is illustrated for neutron-rich 
Nickel isotopes close to the drip line.
Perspectives for unbound HFB theory and QRPA calculations using the Gamow-HFB quasiparticle basis 
will then be discussed.

\section{General HFB formalism with DFT}
The HFB equations are expressed in super-matrix form 
constituted by particle-hole field Hamiltonian $h$,
particle-particle pairing Hamiltonian $\tilde{h}$ 
and chemical potential $\lambda$ 
guaranteeing conservation of particle number in average:
\begin{eqnarray}&&\left(
\begin{array} {cc}
h - \lambda & \tilde{h} \\
\tilde{h} & \lambda - h
\end{array}
\right)
\left(
\begin{array} {c}
u \\ v 
\end{array}
\right)
=
E
\left(
\begin{array} {c}
u \\ v 
\end{array}
\right). \label{HFB_matrix}
\end{eqnarray}
Using Skyrme and density-dependent contact interactions 
for the particle-hole and pairing channels, respectively, 
$h$ and $\tilde{h}$ are expressed in terms of local normal density 
$\rho(r)$ and pairing density $\tilde{\rho}(r)$.
Formulas providing $\rho$, $\tilde{\rho}$, $h$ and $\tilde{h}$ 
can be found in \cite{Dob84,Karim_code}. 
As $h$ and $\tilde{h}$ depend on $\rho$ and $\tilde{\rho}$, 
determined from quasiparticles eigenvectors of Eq.~(\ref{HFB_matrix}),
the HFB equations must be solved in a self-consistent manner 
\cite{Ring_Schuck}.

Let us consider the HFB equations with the Skyrme energy density functionals 
and density-dependent contact pairing interactions 
assuming spherical symmetry.
Fixing orbital and total angular momentum $\ell$ and $j$, 
as well as proton or neutron nature of the wave functions,
Eq.~(\ref{HFB_matrix}) becomes a system of radial differential 
equations \cite{Dob84}:
\begin{eqnarray}
&&\left( \frac{d}{dr} \frac{\hbar^2}{2 m^*(r)} \frac{d}{dr} \right) u(k,r)
= \left[ \frac{\hbar^2 \ell(\ell+1)}{2 m^*(r) r^2} + V(r) - (\lambda + E) \right] u(k,r) + W(r)~v(k,r), \nonumber \\
&&\left( \frac{d}{dr} \frac{\hbar^2}{2 m^*(r)} \frac{d}{dr} \right) v(k,r)
= \left[ \frac{\hbar^2 \ell(\ell+1)}{2 m^*(r) r^2} + V(r) - (\lambda - E) \right] v(k,r) - W(r)~u(k,r), \nonumber \\ \label{HFB_eq}
\end{eqnarray}
where 
\begin{itemize}
\item $u(k,r)$ and $v(k,r)$ are respectively the upper and lower components of 
quasiparticle wave function with energy $E$, 
and $k = \sqrt{2mE}/{\hbar}$ with the nucleon mass $m$,
\item $m^*(r)$, $V(r)$ and $W(r)$ are respectively the effective mass, 
the particle-hole (field) and particle-particle (pairing) potentials of 
the HFB Hamiltonian. 
\end{itemize}
Because nuclear interactions are finite range, only Coulomb and 
centrifugal parts remain for $r \rightarrow +\infty$,
so that Eq.~(\ref{HFB_eq}) becomes asymptotically:
\begin{eqnarray}
\frac{d^2 u}{dr^2}(k,r) &=& \left( \frac{\ell(\ell+1)}{r^2} + \frac{2 \eta_u k_u}{r} - k_u^2 \right) u(k,r), \nonumber \\
\frac{d^2 v}{dr^2}(k,r) &=& \left( \frac{\ell(\ell+1)}{r^2} + \frac{2 \eta_v k_v}{r} - k_v^2 \right) v(k,r), \label{HFB_asymp_eq}
\end{eqnarray}
where the generalized momenta $k_u,k_v$ and their associated Sommerfeld 
parameters $\eta_u,\eta_v$ are defined by 
\begin{eqnarray}
k_u &=& \sqrt{\frac{2m}{\hbar^2}(\lambda+E)} \mbox{ , } k_v = 
\sqrt{\frac{2m}{\hbar^2}(\lambda-E)}, \label{ku_kv_def} \\
\eta_{u(v)} &=& \frac{m Z C_c}{\hbar^2 k_{u(v)}} \mbox{ (proton)} \mbox{ , } 
\eta_{u(v)} = 0 \mbox{ (neutron)} \label{eta_u_eta_v_def}
\end{eqnarray}
with the number of protons $Z$ and the Coulomb constant $C_c$.
Hence, $u(k,r)$ and $v(k,r)$ are linear combinations of the Hankel or 
Coulomb wave functions $H^{\pm}_{\ell \eta_{u(v)}} (k_{u(v)} r)$ 
for $r \rightarrow + \infty$.
Note that $k_v$ is always imaginary provided the HFB ground state is bound 
($\lambda < 0$),
while $k_u$ is real (imaginary) for $E > - \lambda$ ($E < - \lambda$).

The chemical potentials $\lambda$ for neutrons and protons are determined 
from the requirement of conservation of their number in average:
\begin{eqnarray}
\langle \hat{N} \rangle = \sum_{i} N_i = N \mbox{ , } N_i = \int_{0}^{+\infty} v_i^2(r)~dr, \label{av_condition}
\end{eqnarray}
(and similar equations for protons). 
Here the sum runs over all single-particle states, 
$N$ is the number of neutrons and $\langle \hat{N} \rangle$ is 
the expectation value in the HFB ground state.
For a given particle-hole field Hamiltonian $h$, 
the chemical potential $\lambda$ could be calculated in principle 
exactly at each iteration, 
recalculating all quasiparticle wave functions 
from Eq.~(\ref{HFB_matrix}) and updating $\lambda$ 
until Eq.~(\ref{av_condition}) is verified. 
However, in practice, it is much faster to use instead an approximate
chemical potential issued from the BCS formulas, 
which will converge self-consistently to the exact chemical potential 
along with the HFB Hamiltonian 
\cite{Dob84}. For that, one defines auxiliary single-particle energies 
$\bar{e}_i$ and auxiliary pairing gaps $\bar{\Delta}$ by 
\begin{eqnarray}
\bar{e}_i &=& \lambda + E_i (1 - 2 N_i) \mbox{ , } \bar{\Delta}_i 
= 2 E_i \sqrt{N_i (1-N_i)}, \label{eq_e_Delta}
\end{eqnarray}
which are defined by applying the BCS type formula 
to the HFB quasiparticle energies 
$E_i$, the average particle number $N_i$ defined in Eq.~(\ref{av_condition}) 
and the chemical potential $\lambda$ issued from the previous iteration.
While $\bar{e}_i$ and $\bar{\Delta}_i$ correspond to the single-particle 
energy and the pairing gap in the BCS approximation, 
they are used here as auxiliary variables to solve the HFB equations. 
The approximate chemical potential $\lambda$ is obtained 
by solving its associated BCS equation:
\begin{eqnarray}
\sum_{i} \left( 1 - \frac{\bar{e}_i - \lambda}{\sqrt{(\bar{e}_i - \lambda)^2 + \bar{\Delta}_i^2}} \right) = 2 N. \label{lambda_eq}
\end{eqnarray}

\section{S-matrix poles and scattering quasiparticle states}
\subsection{Boundary conditions}
The upper and lower components, $u(k,r)$ and $v(k,r)$, of the quasiparticle 
wave function satisfy the following boundary conditions:
\begin{eqnarray}
u(k,r) &\sim& C_u^0 r^{\ell+1} \mbox{ , } v(k,r) \sim C_v^0 r^{\ell+1} \mbox{ , } r \rightarrow 0 \label{boundary_zero} \\
u(k,r) &\sim& C_u^+ H^+_{\ell \eta_u}(k_u r) + C_u^- H^-_{\ell \eta_u}(k_u r) \mbox{ , } r \rightarrow +\infty  \label{u_boundary_inf} \\
v(k,r) &\sim& C_v^+ H^+_{\ell \eta_u}(k_v r)  \mbox{ , } r \rightarrow +\infty.  \label{v_boundary_inf}
\end{eqnarray}
Eq.~(\ref{boundary_zero}) is required by regularity of wave functions at $r=0$.
Eqs.~(\ref{u_boundary_inf}) and (\ref{v_boundary_inf}) determine 
the nature of quasiparticle state, which can be a bound, 
resonant $(C_u^- = 0)$ or scattering $(C_u^- \neq 0)$ state, 
and are generalizations of the boundary conditions defining single-particle 
states using the Berggren completeness relation.
Eq.~(\ref{v_boundary_inf}) demands outgoing wave function behavior of $v(r)$ 
for all quasiparticle states. 
If its energy $E$ is real and positive, as in the standard HFB approach, 
Eq.~(\ref{v_boundary_inf}) is equivalent to the asymptotic condition 
$v(k,r) \rightarrow 0$ for $r \rightarrow +\infty$; 
the condition arising from integrability of nuclear density over all space 
\cite{Dob84}. Extension to complex energies follows from
analyticity of the $v(k,r)$ function in the complex $k$-plane. 
Eq.~(\ref{u_boundary_inf}) with $C_u^- = 0$ then defines 
quasiparticle S-matrix poles, as it
is equivalent to $u(k,r) \rightarrow 0$ for $r \rightarrow +\infty$ 
for bound quasiparticle states with $E < |\lambda|$, 
and provides resonant quasiparticle states if $E$ is complex. 
Eq.~(\ref{u_boundary_inf}) with $C_u^- \neq 0$ 
represents standard scattering quasiparticle states for real and positive $E$, 
but they are extended to complex energies by analyticity arguments.

\subsection{Normalization of quasiparticle states} \label{norm_qp}
Bound HFB quasiparticle states with energy $E_n$ are normalized by:
\begin{eqnarray}
\int_{0}^{+\infty} \left[ u(k_n,r)^2 + v(k_n,r)^2 \right]~dr = 1. \label{qp_bound_norm}
\end{eqnarray}
where $k_n = \sqrt{2mE_n}/\hbar$.
For resonant quasiparticle states, the integral in the above equation diverges, 
so that this normalization condition cannot be used.
The complex scaling method has been known as a practical means to normalize 
single-particle resonance states \cite{Vertse_CS}.
Convergence of integrals is obtained therein integrating up to a finite 
radius $R$ situated in the asymptotic region, 
after which the interval $[R:+\infty[$ is replaced by a complex contour 
defined by a rotation angle $\theta > 0$,
allowing exponential decrease of the integrand. 
Owing to Eqs.~(\ref{u_boundary_inf}) and (\ref{v_boundary_inf}),
the same method can be used to normalize resonant quasiparticle states, 
so that Eq.~(\ref{qp_bound_norm}) becomes:
\begin{eqnarray}
&&\int_{0}^{R} \left[ u(k_n,r)^2 + v(k_n,r)^2 \right]~dr \nonumber \\
&+& \int_{0}^{+\infty} \left[ C_u^+ H^+_{\ell \eta_u}(k_u (R + x e^{i\theta_u})) \right]^{2} e^{i\theta_u}~dx
+ \int_{0}^{+\infty} \left[ C_v^+ H^+_{\ell \eta_v}(k_v (R + x e^{i\theta_v})) \right]^{2} e^{i\theta_v}~dx
= 1, \label{qp_complex_scaling_norm}
\end{eqnarray}
where $\theta_u > 0$ and $\theta_v > 0$ are chosen such that 
improper integrals converge. 
Hence, as in the single-particle case, 
normalization of quasiparticle S-matrix poles presents no other difficulty.
As in Ref.~\cite{PRC1_michel}, complex-scaled integrals will be denoted 
$\displaystyle Reg \left[ \int_{0}^{+\infty} f(r)~dr \right]$,
i.e.~the regularized value of the diverging integral.

Scattering quasiparticle states must be orthonormalized 
with the Dirac delta distribution:
\begin{eqnarray}
\int_{0}^{+\infty} \left[ u(k_a,r) u(k_b,r)~dr + v(k_a,r) v(k_b,r) \right]~dr = \delta(k_a - k_b), \label{Dirac_delta_norm}
\end{eqnarray}
for those with momenta $k_a$ and $k_b$.
From Eqs.~(\ref{u_boundary_inf}) and (\ref{v_boundary_inf}), 
assuming that Eq.~(\ref{HFB_asymp_eq}) is obtained for $r \geq R$,
Eq.~(\ref{Dirac_delta_norm}) becomes:
\begin{eqnarray}
&&\int_{0}^{R} \left[ u(k_a,r) u(k_b,r) + v(k_a,r) v(k_b,r) \right]~dr  \nonumber \\
&+& C_{u_a}^+ C_{u_b}^+ Reg \left[ \int_{R}^{+\infty} H^+_{\ell \eta_{u_a}}(k_{u_a} r) H^+_{\ell \eta_{u_b}}(k_{u_b} r)~dr \right]
  + C_{u_a}^- C_{u_b}^- Reg \left[ \int_{R}^{+\infty} H^-_{\ell \eta_{u_a}}(k_{u_a} r) H^-_{\ell \eta_{u_b}}(k_{u_b} r)~dr \right] \nonumber \\
&+& C_{v_a}^+ C_{v_b}^+ Reg \left[ \int_{R}^{+\infty} H^+_{\ell \eta_{v_a}}(k_{v_a} r) H^+_{\ell \eta_{v_b}}(k_{v_b} r)~dr \right] \nonumber \\
&+& C_{u_a}^- C_{u_b}^+ \int_{R}^{+\infty} H^-_{\ell \eta_{u_a}}(k_{u_a} r) H^+_{\ell \eta_{u_b}}(k_{u_b} r)~dr
  + C_{u_a}^+ C_{u_b}^- \int_{R}^{+\infty} H^+_{\ell \eta_{u_a}}(k_{u_a} r) H^-_{\ell \eta_{u_b}}(k_{u_b} r)~dr \nonumber \\
&=& \delta(k_a - k_b). \label{Dirac_delta_norm_exp}
\end{eqnarray}
The divergence of the Dirac delta function at $k_a = k_b$ occurs 
by way of the two last integrals of Eq.~(\ref{Dirac_delta_norm_exp}), 
as no complex scaling can make them converge if $k_a = k_b$ \cite{PRC1_michel}. 
The Dirac delta normalization of the Coulomb wave functions implies, 
as in the single-particle case:
\begin{eqnarray}
&& C_{u_a}^- C_{u_b}^+ \int_{R}^{+\infty} H^-_{\ell \eta_{u_a}}(k_{u_a} r) H^+_{\ell \eta_{u_b}}(k_{u_b} r)~dr
 + C_{u_a}^+ C_{u_b}^- \int_{R}^{+\infty} H^+_{\ell \eta_{u_a}}(k_{u_a} r) H^-_{\ell \eta_{u_b}}(k_{u_b} r)~dr \nonumber \\
&=& 2 \pi C_{u_a}^+ C_{u_a}^- \delta(k_{u_a} - k_{u_b}) + f(k_{u_a},k_{u_b}), \label{partial_Dirac_delta}
\end{eqnarray}
where $f(k_{u_a},k_{u_b})$ is finite for all $(k_{u_a},k_{u_b})$.
The relation between $\delta(k_a - k_b)$ and $\delta(k_{u_a} - k_{u_b})$ 
is easily obtained from Eq.~(\ref{ku_kv_def}):
\begin{eqnarray}
&&\delta(k_{u_a} - k_{u_b}) = \left[ \frac{\partial k_{u_a}}{\partial k_a}(k_a) \right]^{-1} \delta(k_a - k_b) = \frac{k_{u_a}}{k_a} \delta(k_a - k_b).
\label{delta_eq}
\end{eqnarray}
This a direct application of the standard Dirac delta distribution property 
stating that $\displaystyle \delta(f(k)) = f'(k_0)^{-1} \delta(k-k_0)$
for a given function $f(k)$ bearing a unique simple zero at $k = k_0$ 
\cite{Messiah}. 
Note that $k_b$ is fixed while $k_a$ is varied to obtain Eq.~(\ref{delta_eq}).
Inserting Eqs.~(\ref{partial_Dirac_delta}) and (\ref{delta_eq}) 
into Eq.~(\ref{Dirac_delta_norm_exp}), one obtains:
\begin{eqnarray}
&&\int_{0}^{+\infty} \left[ u(k_a,r) u(k_b,r)~dr + v(k_a,r) v(k_b,r) \right]~dr = \delta(k_a - k_b) \nonumber \\
&\Leftrightarrow& \frac{2 \pi k_{u_a}}{k_a} C_{u_a}^+ C_{u_a}^- \delta(k_a - k_b) =\delta(k_a - k_b) + g(k_a,k_b), \label{Dirac_delta_eq}
\end{eqnarray}
where $g(k_a,k_b)$ bears the same properties as $f(k_{u_a},k_{u_b})$.
As quasiparticle scattering states are orthogonal for $k_a \neq k_b$, 
$g(k_a,k_b) = 0$ therein,
so that $\delta(k_a - k_b) + g(k_a,k_b) = \delta(k_a - k_b)$ in all cases. 

Dirac delta distribution normalization for scattering states 
$| k \rangle$ and $| k' \rangle$ immediately follows:
\begin{eqnarray}
\langle k | k' \rangle = \delta(k - k') \Leftrightarrow C_u^+ C_u^- = \frac{k}{2 \pi k_u}. \label{Dirac_delta_norm_eq}
\end{eqnarray}
Hence, besides the additional factor $k/k_u$, 
the normalization condition for quasiparticle scattering states is 
the same as that for single-particle scattering states \cite{PRC1_michel}.

\subsection{Completeness of quasiparticle states of real and complex energy}
The HFB supermatrix defined in Eq.(\ref{HFB_matrix}) is hermitian, so that it possesses a spectral decomposition
\cite{Dunford_Schwarz}:
\begin{eqnarray}
\sum_{n \in b} \left[ u(k_n,r) u(k_n,r') + v(k_n,r) v(k_n,r') \right] 
+ \int_{k_{\lambda}}^{+\infty} \left[ u(k,r) u(k,r') + v(k,r) v(k,r') \right]~dk = \delta(r - r') \label{qp_comp},
\end{eqnarray}
where $k_n=\sqrt{2mE_n}/\hbar$ for a bound quasisparticle state 
with energy $E_n$, 
$k$ is a linear momentum for a continuum quasiparticle state,
$u(\kappa,r), v(\kappa,r)$ ($\kappa = k_n$ or $k$) are respectively 
the upper and lower components of a quasiparticle wave function 
with quantum numbers $\ell$ and $j$ (here implicit), and $k_{\lambda} = \sqrt{-2m \lambda}/\hbar$.
All quasiparticle states must be normalized to one (bound) or to a Dirac delta (scattering) (see Sec.(\ref{norm_qp})).
Eq.(\ref{qp_comp}) can also be demonstrated extending the method of Ref.~\cite{JMP_Michel} to quasi-particle states.

In order to obtain Berggren completeness of quasiparticle states, one can proceed as in Ref.~\cite{Lind}, deforming 
the real energy contour in the complex plane. Resonant quasiparticle states appear therein, due to the Cauchy theorem,
as S-matrix poles \cite{Lind}. Hence, Eq.~(\ref{qp_comp}) becomes after contour deformation:
\begin{eqnarray}
\sum_{n \in (b,d)} \left[ u(k_n,r) u(k_n,r') + v(k_n,r) v(k_n,r') \right] 
+ \int_{L^+}^{+\infty} \left[ u(k,r) u(k,r') + v(k,r) v(k,r') \right]~dk = \delta(r - r') \label{Berggren_qp_comp},
\end{eqnarray}
where $k_n$ refers now to a bound $(b)$ or resonant $(d)$ (decaying) 
quasiparticle state and $k$ is complex as it follows the deformed contour in the complex plane, denoted as $L^+$. 
Resonant quasiparticle states are normalized with complex scaling (see Sec.(\ref{norm_qp})).

\section{Numerical determination of quasiparticle energies and wave functions 
with direct integration}
\subsection{Quasiparticle Jost functions}
In Eqs.~(\ref{boundary_zero}), (\ref{u_boundary_inf}) and 
(\ref{v_boundary_inf}), 
constants and momenta of S-matrix poles are determined by 
the requirement of continuity of both the $u(k,r)$ and $v(k,r)$ functions 
and associated derivatives.
These conditions can be expressed in a form of quasiparticle Jost functions, 
defined as a generalization of the Jost function for single-particle problems, 
whose zeros correspond to S-matrix poles \cite{Newton_book}. They read:
\begin{eqnarray}
J_u \left(k,\frac{C_v^0}{C_u^0},\frac{C_v^+}{C_u^+} \right) &=& \frac{u'(k,R_0^+)}{u(k,R_0^+)} - \frac{u'(k,R_0^-)}{u(k,R_0^-)}, \nonumber \\
J_v \left(k,\frac{C_v^0}{C_u^0},\frac{C_v^+}{C_u^+} \right) &=& \frac{v'(k,R_0^+)}{v(k,R_0^+)} - \frac{v'(k,R_0^-)}{v(k,R_0^-)}, \nonumber \\
J_m \left(k,\frac{C_v^0}{C_u^0},\frac{C_v^+}{C_u^+} \right) &=& \frac{u(k,R_0^+)}{u(k,R_0^-)} - \frac{v(k,R_0^+)}{v(k,R_0^-)}, \label{quasi_part_Jost_pole}
\end{eqnarray}
where $R_0$ is a radius typically chosen around the nuclear surface and 
one can demand arbitrarily that $C_u^0 = C_u^+ = 1$ 
in Eqs.~(\ref{boundary_zero}) and (\ref{u_boundary_inf}). 
The functions, $u(k,R_0^+)$, $v(k,R_0^+)$ and their derivatives, 
are obtained by forward integration of Eq.~(\ref{HFB_eq}) 
using Eq.~(\ref{boundary_zero}) as initial conditions, 
while $u(k,R_0^-)$, $v(k,R_0^-)$ and their derivatives
are calculated by backward integration of Eq.~(\ref{HFB_eq}) 
from the initial conditions provided by 
Eqs.~(\ref{u_boundary_inf}) and (\ref{v_boundary_inf}).
In Eq.~(\ref{quasi_part_Jost_pole}), one can clearly see that 
$u(k,r)$ and $v(k,r)$ will have continuous logarithmic derivatives 
if $J_u = 0$ and $J_v = 0$ respectively. 
However, these two equalities are not sufficient to uniquely determine 
the quasiparticle state. Indeed, they imply that 
one can choose a set of constants so that either $u(k,r),u'(k,r)$, 
or $v(k,r),v'(k,r)$ are continuous, 
but not necessarily both of them.
The condition $J_m = 0$ is thus enforced in Eq.~(\ref{quasi_part_Jost_pole}). 
The set of three equations, $J_u = 0$, $J_v = 0$ and $J_m = 0$, 
uniquely determine quasiparticle S-matrix poles.

For quasiparticle scattering states, the linear momentum $k$ is fixed, 
but constants have to be calculated with a matching procedure.
One starts with imposing the condition $C_u^0 = 1$, as for S-matrix poles.
As the $u(k,r)$ component is of scattering type, the condition $J_u = 0$ 
can always be fulfilled with appropriately chosen $C_u^+$ and $C_u^-$ constants.
Thus, it is sufficient to deal only with $J_v$ and $J_m$:
\begin{eqnarray}
J_v \left(\frac{C_v^0}{C_u^0},C_v^+ \right) &=& \frac{v'(k,R_0^+)}{v(k,R_0^+)} - \frac{v'(k,R_0^-)}{v(k,R_0^-)}, \nonumber \\
J_m \left(\frac{C_v^0}{C_u^0},C_v^+ \right) &=& \frac{u(k,R_0^+)}{u(k,R_0^-)} - \frac{v(k,R_0^+)}{v(k,R_0^-)}, \label{quasi_part_Jost_scat}
\end{eqnarray}
the difference with Eq.~(\ref{quasi_part_Jost_pole}) being that 
$J_v$ and $J_m$ now depend on two parameters instead of three.
As in the S-matrix pole equations, $u(k,R_0^+)$, $v(k,R_0^+)$ and 
their derivatives are generated by forward integration of Eq.~(\ref{HFB_eq}).
Concerning the implementation of $u(k,R_0^-)$, $v(k,R_0^-)$ and 
their derivatives, however, one first continues integrating forward 
in order to obtain $u(k,R),u'(k,R)$, $R$ being in the asymptotic region. 
At this point $R$, $u(k,R),u'(k,R)$ provide an initial condition for 
backward integration, 
while Eq.~(\ref{v_boundary_inf}) is used to initialize $v(k,R),v'(k,R)$. 
In this way, we obtain $u(k,R_0^-)$, $v(k,R_0^-)$ and their derivatives.
Thus, the equations $J_v = 0$ and $J_m = 0$ provide the matching constants 
rendering $v(k,r), v'(k,r)$ continuous.

The conditions, $J_u = 0$ (for S-matrix poles), $J_v = 0$ and $J_m = 0$, 
form a system of non-linear equations of two or three dimensions. 
Consequently, it has to be solved with multi-dimensional Newton method. 
The only problem therein is to find a good starting point from where 
one can attain fast convergence to the exact solution in a stable way.

\subsection{Determination of quasiparticle energy and integration constants}
Following Ref.~\cite{Karim_code}, it is convenient to introduce 
linearly independent solutions of Eq.~(\ref{HFB_eq})
in order to determine the constants defined in 
Eqs.~(\ref{boundary_zero}), (\ref{u_boundary_inf}) and (\ref{v_boundary_inf}):
\begin{eqnarray} 
\left( \begin{array} {c} u \\ v \end{array} \right)
&=& C_u^0 \left( \begin{array} {c} f_{u_0} \\ g_{u_0} \end{array} \right)
+ C_v^0 \left( \begin{array} {c} f_{v_0} \\ g_{v_0} \end{array} \right), 
\label{uv_forward}  \\
\left( \begin{array} {c} u \\ v \end{array} \right) 
&=& C_u^+ \left( \begin{array} {c} f_{u^+} \\ g_{u^+} \end{array} \right)
+ C_u^- \left( \begin{array} {c} f_{u^-} \\ g_{u^-} \end{array} \right) 
+ C_v^+ \left( \begin{array} {c} f_{v^+} \\ g_{v^+} \end{array} \right), \label{uv_backward}
\end{eqnarray}
where the introduced basis functions verify:
\begin{eqnarray}
&&f_{u_0}(r) \sim r^{\ell+1} \mbox{ , } g_{v_0}(r) \sim r^{\ell+1} \mbox{ , } 
f_{v_0}(r) \sim D_0 r^{\ell+3} \mbox{ , } g_{u_0}(r) \sim -D_0 r^{\ell+3} \mbox{ , } r \rightarrow 0, \nonumber \\
&&f_{u^\pm}(r) \sim H^{\pm}_{\ell \eta_u}(k_u r) \mbox{ , } 
g_{v^+}(r) \sim H^+_{\ell \eta_v}(k_v r) \mbox{ , } f_{v^+}(r) \rightarrow 0  \mbox{ , }  
g_{u^\pm}(r) \rightarrow 0 \mbox{ , }r \rightarrow +\infty, \label{fg_boundary_cond}
\end{eqnarray}
with 
\begin{eqnarray}
D_0 = \frac{m^*(0) W(0)}{(2\ell + 3) \hbar^2}. \label{D0}
\end{eqnarray}
Eq.~(\ref{D0}) is obtained inserting $u(r) = r^{\ell +1}$ and 
$v(r) = -D_0 r^{\ell + 3}$ in the second equality of Eq.~(\ref{HFB_eq})
and solving the equation keeping only dominant terms.

As the basis functions of Eqs.~(\ref{uv_forward}) and (\ref{uv_backward}) 
depend only on $k$ of the quasiparticle state,
they can be calculated with direct integration, 
in a forward direction for Eq.~(\ref{uv_forward}) and 
in a backward direction for Eq.~(\ref{uv_backward}).
Used methods to determine quasiparticle wave function differ 
according to their characters; 
S-matrix poles or scattering states, as discussed below. 

\subsection{Bound and resonant quasiparticle states}
To find S-matrix poles, it is first necessary to start with 
a good approximation of $k$, denoted $k_{app}$. For that,
a no-pairing approximation is firstly performed. 
Neglecting $\tilde{h}$ in Eq.~(\ref{HFB_matrix}), 
the Gamow-HFB equations reduce to the GHF equations:
\begin{eqnarray}
h | \phi_i \rangle = e_i |\phi_i \rangle, \label{no_pair_app}
\end{eqnarray}
where $e_i$ are complex (real) for resonant (bound) states.
Eq.~(\ref{no_pair_app}) provides bound and narrow resonant single-particle 
states of interest, which will be in finite number.
As pairing potential $\tilde{h}$ is weak compared to $h$,
there will always be unique correspondence between 
the GHF single-particle S-matrix poles and the HFB quasiparticle S-matrix poles.
Unless the quasiparticle S-matrix poles lie close to the Fermi energy, 
their lower (upper) components will be very close to 
$\phi_i(r)$ if $| \phi_i \rangle$ are (un)occupied at the HF level,
so that the auxiliary energies $\bar{e}_i$, defined in Eq.~(\ref{eq_e_Delta}), 
will be very close to the real parts of $e_i$.
Secondly, the HFB matrix in Eq.~(\ref{HFB_matrix}) is diagonalized.
It has been found that the use of a P{\"o}schl-Teller-Ginocchio (PTG) basis 
provides sufficiently precise results \cite{HFB_PTG_michel}.
Therefore, for $E_i$ in Eq.~(\ref{eq_e_Delta}) 
we use the quasiparticle energies 
obtained by diagonalizing the HFB matrix in the PTG basis.
For a given GHF state of energy $e_i$, 
the starting quasiparticle energy $E_{app}$ 
(from which $k_{app}$ is immediately deduced), 
is then the BCS quasiparticle energy 
whose $\bar{e}_i$ is closest to the real part of $e_i$. 
If the HFB quasiparticle S-matrix pole is far from the Fermi energy, 
$E_{app}$ is very close to the exact energy.
Otherwise, it will still provide a good starting point, as, in practice,
one can have only one quasiparticle state close to the Fermi energy 
for a given $(\ell,j)$-partial wave.

Furthermore, one demands $C_u^- = 0$, 
which translates into a linear eigen-value problem of dimension equal to four, 
deduced from Eqs.~(\ref{uv_forward}) and (\ref{uv_backward}),
which one matches at $r=R_0$:
\begin{eqnarray}
\left( \begin{array} {cccc}
f_{u_0} & f_{v_0} & -f_{u^+} & -f_{v^+} \\
g_{u_0} & g_{v_0} & -g_{u^+} & -g_{v^+} \\
f'_{u_0} & f'_{v_0} & -f'_{u^+} & -f'_{v^+} \\
g'_{u_0} & g'_{v_0} & -g'_{u^+} & -g'_{v^+}
\end{array} \right)
\left( \begin{array} {c}
C_u^0 \\ C_v^0 \\ C_u^+ \\ C_v^+
\end{array} \right) = 0, \label{Cu_Cv_eigensystem}
\end{eqnarray}
where all matrix functions have been evaluated at $r = R_0$ by way of 
backward or forward integration.
As the integration constants are not simultaneously equal to zero, 
they have to form an eigenvector of the matching matrix of 
Eq.~(\ref{Cu_Cv_eigensystem}), 
which we denote $M$ hereafter, of eigenvalue equal to zero. 
However, the determinant of the $4 \times 4$ matrix $M$ is zero 
uniquely for the exact value of $k$.
Thus, the set of approximate constants to use as a starting point 
for Newton method is chosen as the eigenvector of $^tM M$
whose associated eigenvalue is the smallest in modulus 
($^tM M$ is used instead of $M$ because it is symmetric).
The constant ratios $C_v^0/C_u^0$ and $C_v^+/C_u^+$ used in 
Eq.~(\ref{quasi_part_Jost_pole}) follow, 
as they are independent of the norm of the considered eigenvector. 
Exact determination of $k$, $C_v^0/C_u^0$ and $C_v^+/C_u^+$ can then be 
worked out via three-dimensional Newton method.

\subsection{Scattering quasiparticle state}
If one considers a scattering state, it is convenient to define 
$a^+,a^-,b^+,b^-$ so that $C_u^{\pm} = a^{\pm} C_u^0 + b^{\pm} C_v^0$.
Moreover, as all constants are calculated up to a normalization factor, 
one can impose $C_u^0 = 1$.
Upper components of Eqs.~(\ref{uv_forward}) and (\ref{uv_backward}) 
matched at $r=R$ and Eq.~(\ref{fg_boundary_cond}) 
provide linear equations for $a^{\pm}$ and $b^{\pm}$:
\begin{eqnarray}
&&a^+ f_{u^+}(R) + a^- f_{u^-}(R) = f_{u_0}(R) \mbox{ , }b^+ f_{u^+}(R) + b^- f_{u^-}(R) = f_{v_0}(R), \nonumber \\
&&a^+ f'_{u^+}(R) + a^- f'_{u^-}(R) = f'_{u_0}(R) \mbox{ , } b^+ f'_{u^+}(R) + b^- f'_{u^-}(R) = f'_{v_0}(R). \label{ab_linear_system}
\end{eqnarray}
From the knowledge of $a^{\pm}$ and $b^{\pm}$, 
matching lower components in Eqs.~(\ref{uv_forward}) and (\ref{uv_backward}) 
at $r = R_0$ determines $C_v^0$ and $C_v^+$ via linear equations as well:
\begin{eqnarray}
&&C_v^0 [g_{v_0}(R_0) - b^+ g_{u^+}(R_0) - b^- g_{u^-}(R_0)] - C_v^+ g_{v^+}(R_0) = a^+ g_{u^+}(R_0) + a^- g_{u^-}(R_0) - g_{u_0}(R_0), \nonumber \\ 
&&C_v^0 [g_{v_0}'(R_0) - b^+ g'_{u^+}(R_0) - b^- g'_{u^-}(R_0)] - C_v^+ g'_{v^+}(R_0) = a^+ g'_{u^+}(R_0) + a^- g'_{u^-}(R_0) - g'_{u_0}(R_0). 
\label{Cv_linear_system}
\end{eqnarray}
As $C_u^{\pm} = a^{\pm} + b^{\pm} C_v^0$,
all constants are determined with simple two-dimensional linear systems.
Newton method applied to Eq.~(\ref{quasi_part_Jost_scat}) converges 
very quickly using the obtained set of constants as a starting point.
Note that the use of $H^{\pm}_{\ell \eta_u}(k_u r)$ functions in 
Eqs.~(\ref{uv_backward}) and (\ref{fg_boundary_cond}) can be sometimes 
unstable, especially for the proton case, where, for low scattering energies, 
they can be very large and cancel almost exactly in Eq.~(\ref{u_boundary_inf}).
In this case, it is better to use regular and irregular Coulomb wave functions, 
$F_{\ell \eta_u}(k_u r)$ and $G_{\ell \eta_u}(k_u r)$, 
as basis functions.

\section{Normal and Pairing Densities} \label{dens_def}

\begin{figure}[htb]
\includegraphics[angle=-90,width=1\textwidth]{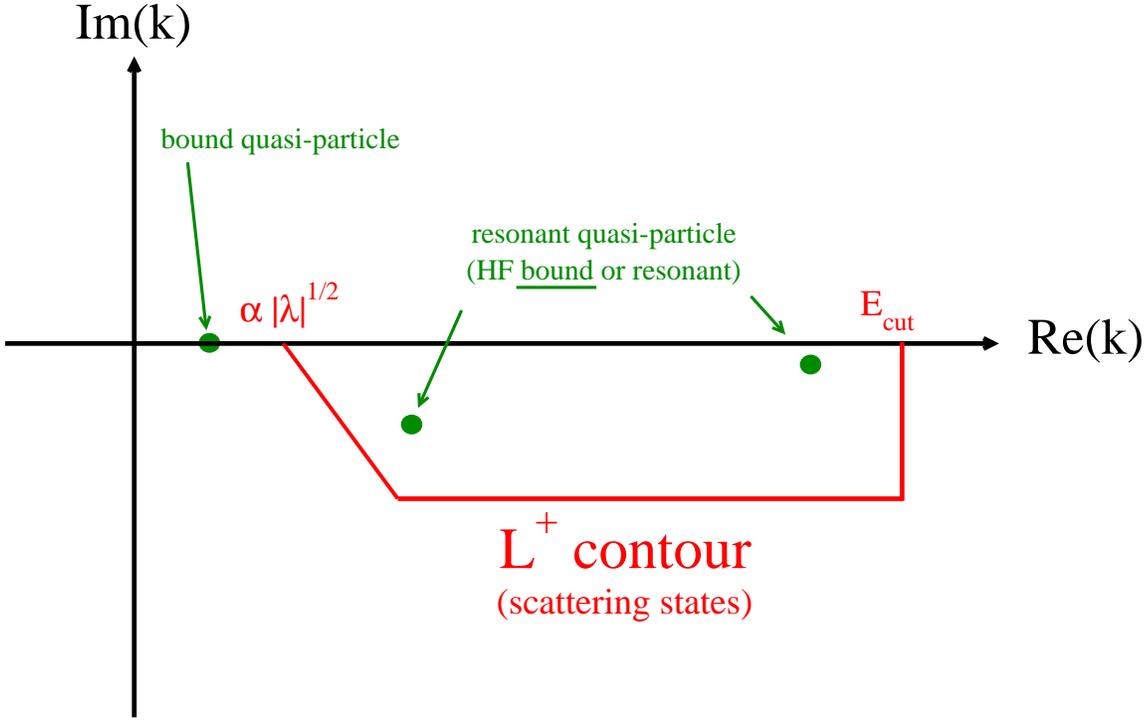}
\protect\caption{\label{Lplus_quasi_part}  (color online) 
Location of quasiparticle S-matrix poles and 
deformed complex contour $L^+$ of scattering quasiparticle states 
used in the Berggren completeness relation.
Here, $\alpha = \sqrt{2m}/\hbar$.}
\end{figure}

As quasiparticle states of complex energy form a complete set (see Eq.(\ref{Berggren_qp_comp})), 
one can directly express densities with upper and lower components of quasiparticle states:
\begin{eqnarray}
\rho_{\ell j}(r) &=& \sum_{n \in (b,d)} v^2(k_n,r) + \int_{L^+} \!\!\!\! v^2(k,r)~dk  \mbox{ , } \rho(r) = \sum_{\ell j} \rho_{\ell j}(r), \nonumber \\
\tilde{\rho}_{\ell j}(r) &=& -\sum_{n \in (b,d)} u(k_n,r) v(k_n,r) - \int_{L^+} \!\!\!\! u(k,r) v(k,r)~dk \mbox{ , }
\tilde{\rho}(r) = \sum_{\ell j} \tilde{\rho}_{\ell j}(r), \label{densities_Gamow_HFB}
\end{eqnarray}
where $\rho_{\ell j}(r)$ and $\tilde{\rho}_{\ell j}(r)$ are respectively 
partial normal and pairing densities related to a given partial wave 
with quantum numbers $\ell$ and $j$, and $\rho(r)$, $\tilde{\rho}(r)$ are 
respectively the normal and pairing densities of the HFB ground state.
%where $k_n$ refers to a bound $(b)$ or resonant $(d)$ (decaying) 
%quasiparticle state and $k$ is complex as it follows the $L^+$ contour of 
%Fig.~\ref{Lplus_quasi_part}. 
However, due to the zero-range character of Skyrme forces, it is necessary
to introduce an energy cut in contour integrals, so that $L^+$ contour has to stop
at finite energy $E_{cut}$ (see Fig.~\ref{Lplus_quasi_part}). 
Note that, due to this requirement, it is necessary for $L^+$ complex contours to come back to the real axis.
Even though quasiparticle wave functions are complex 
in Eq.~(\ref{densities_Gamow_HFB}),
$\rho_{\ell j}(r)$ and $\tilde{\rho}_{\ell j}(r)$ are real 
because one is considering a HFB bound ground state, so that, due to Cauchy theorem,
complex integration in Eq.(\ref{densities_Gamow_HFB}) is equivalent
to real integration in the standard case.
%because Eqs.~(\ref{densities_standard_HFB}) and (\ref{densities_Gamow_HFB}) 
%provide the same result. 
As a consequence, the DFT can be applied also to the Gamow HFB formalism, 
i.e.~potentials $V(r)$ and $W(r)$ in Eq.~(\ref{HFB_eq}) are evaluated 
using the standard formulas of Ref.~\cite{Dob84}. 
As shown in Fig.~\ref{Lplus_quasi_part}, 
the bound HF single-particle states can become
resonant states when pairing correlations are added \cite{giai}. 
Thus, physical interpretation of a resonant quasiparticle is 
somewhat different from that of single-particle resonances: 
widths of the quasiparticle states associated with 
the HF bound single-particle states originates from 
pairing-induced couplings between the bound and scattering states 
\cite{Bel87}.

In the same way as in the Gamow Shell Model 
\cite{PRL_michel,PRC1_michel,PRC2_michel}, 
the scattering $L^+$ contours in Eq.~(\ref{densities_Gamow_HFB}) 
have to be discretized, providing a finite set of linear momenta 
and weights $(k_i,w_i)$.
In practice, the Gauss-Legendre quadrature has been found to be 
most efficient. Scattering quasiparticle states are also renormalized, 
multiplying them by $\sqrt{w_i}$ \cite{Lind},
so that the discretized expressions of Eq.~(\ref{densities_Gamow_HFB}) 
are formally identical to the discrete case:
\begin{eqnarray}
\rho_{\ell j}(r) &\simeq& \sum_{n \in (b,d)} v^2(k_n,r) + \sum_{i} v_{w_i}^2(k_i,r), \nonumber \\
\tilde{\rho}_{\ell j}(r) &\simeq& -\sum_{n \in (b,d)} u(k_n,r) v(k_n,r) - \sum_{i} u_{w_i}(k_i,r) v_{w_i}(k_i,r), \label{discretized_contours}
\end{eqnarray}
where $u_{w_i}(k_i,r) = \sqrt{w_i}~u(k_i,r)$ and $v_{w_i}(k_i,r) = \sqrt{w_i}~v(k_i,r)$.

\section{Another method: expansion of quasiparticle states with the GHF basis} 
\label{two_basis_chapter}
Another possibility to solve the HFB equations in complex energy plane is 
to use the Gamow single-particle states as a basis.
The optimal Berggren basis to expand the HFB quasiparticle states is 
obviously the GHF basis generated by the potential $V(r)$
and the effective mass $m^*(r)$ of Eq.~(\ref{HFB_eq}). 
Note that it is not equivalent to the GHF basis issued from 
the pure HF variational principle in that pairing correlations 
always give extra contributions to the particle-hole part
of the HFB Hamiltonian. 
Indeed, we noticed in our numerical calculation that 
other Berggren bases make the HFB self-consistent procedure unstable 
due to the appearance of very large matrix elements 
in the HFB Hamiltonian matrix. 
The use of the optimized Berggren basis mentioned above removes this problem. 
This approach may be regarded as a generalization of the two-basis method 
\cite{Gal94, Ter97a, Yam01}.

The GHF basis states $\phi(r)$ are defined by the following equation:
\begin{eqnarray}
\left( \frac{d}{dr} \frac{\hbar^2}{2 m^*(r)} \frac{d}{dr} \right) \phi(r)
= \left[ \frac{\hbar^2 \ell(\ell+1)}{2 m^*(r) r^2} + V(r) - e \right] \phi(r), \label{GHF_eq}
\end{eqnarray}
issued directly from Eqs.~(\ref{HFB_eq}) and (\ref{no_pair_app}),
where $e$ is the complex energy of the GHF state.
The HFB Hamiltonian matrix represented with this basis becomes:
\begin{eqnarray}
\!\!\!\!\!\!\!\!\!\!\!\!\!\!\!&&\left(
\begin{array} {cc}
h - \lambda & \tilde{h} \\
\tilde{h} & \lambda - h
\end{array}
\right) 
=
\left( 
\begin{array} {c|c}
\begin{array} {ccc}
 e_1 - \lambda  &         & 0 \\
              & \ddots  &   \\
       0      &         & e_N - \lambda \\ 
\end{array}
 & \tilde{h} \\ \hline 
\tilde{h} &
\begin{array} {ccc}
 \lambda - e_1  &         & 0 \\
              & \ddots  &   \\
       0      &         & \lambda - e_N \\ 
\end{array}
\end{array}
\right), \label{Two_basis_matrix}
\end{eqnarray}
where the continuous Berggren basis is discretized with 
the Gauss-Legendre quadrature (see Sec.~\ref{dens_def})
so that total number of basis states is $N$. 
Its particle-hole part is evidently diagonal and 
matrix elements of $\tilde{h}$ read:
\begin{eqnarray}
\langle \phi_a | \tilde{h} | \phi_b \rangle = \int_{0}^{+\infty} \phi_a(r) W(r) \phi_b(r)~dr, \label{pairing_ME}
\end{eqnarray}
where $| \phi_b \rangle$ and $| \phi_a \rangle$ are the GHF basis states 
and $W(r)$ is the HFB pairing potential defined in Eq.~(\ref{HFB_eq}).
For bound HFB ground states, $W(r)$ decreases sufficiently quickly 
so that no complex scaling is needed to evaluate the integral of 
Eq.~(\ref{pairing_ME}).
Hence, after discretization of the contours representing scattering basis 
states, this method takes a formally identical form to the standard matrix 
diagonalization treatment of the HFB problem.

\section{Numerical applications}
The frameworks described above, 
i.e.~the Gamow-HFB approach in the coordinate or the GHF configurational space, 
are applied to Nickel isotopes close to the neutron drip-line, 
from $^{84}$Ni to $^{90}$Ni, which possess spherical HFB ground states.
In the numerical calculation, the SLy4 Skyrme force \cite{Cha98} is used 
in combination with the surface-type contact pairing interaction 
\cite{Karim_code} whose pairing strength is fitted to reproduce 
the pairing gap of $^{120}$Sn. 
Using the standard notation \cite{Karim_code}, 
the pairing interaction parameters read 
$t_0^{'} = -519.9$ MeV fm$^{3}$ for the density-independent part 
and $t_3^{'} = -37.5t_0^{'}$ MeV fm$^{6}$ for the density-dependent part.
The maximal angular momentum used is $\ell_{max} = 10$ and 
a sharp cut-off at $E_{cut} = 60$ MeV is adopted. 
Scattering contours of quasiparticle states are discretized 
with 60, 100 or 300 Gaussian points. 
Several hundred points are indeed necessary when resonant states lie 
relatively close to $E_{cut}$ (see Fig.~\ref{Lplus_quasi_part}), 
as is the case for the HFB quasiparticle resonance associated with 
the deeply bound neutron $0s_{1/2}$ HF state for example 
(see Table~\ref{neutron_resonant_energies}).
Scattering contours of single-particle states in the GHF basis 
are discretized up to $k_{max} = 4$ fm$^{-1}$ with 100 points, 
which in this case assures convergence of numerical calculation. 
This concerns only for the neutron channel,  
as the pairing gap vanishes in the proton channel.

\begin{figure}[htb]
\includegraphics[angle=-90,width=1\textwidth]{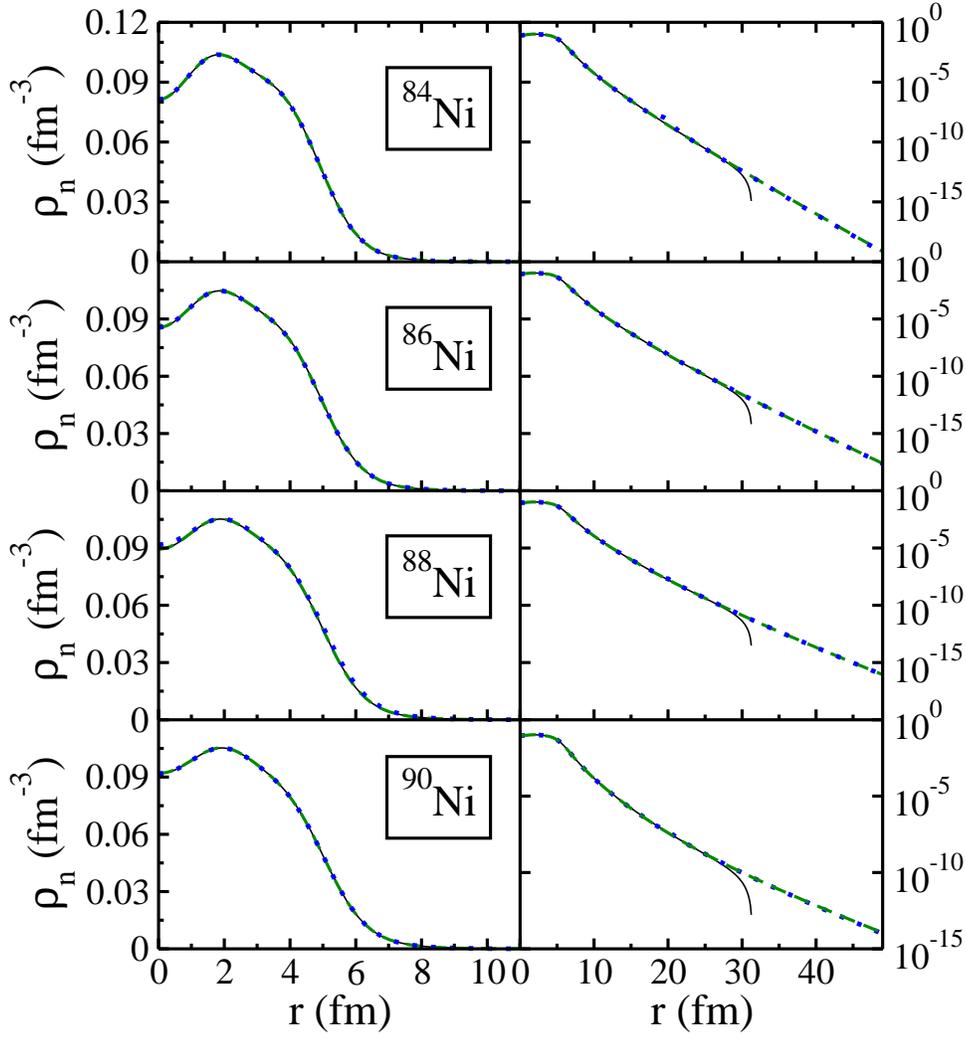}
\protect\caption{\label{dens_Ni}  (color online) 
Neutron densities $\rho_n$
both in normal (left-hand side) and logarithmic (right-hand side) scales.
Results of the HFB/Box, GHFB/Coord.~and GHFB/Config.~calculations 
are displayed by solid, dashed and dotted lines, respectively.}
\end{figure}

\begin{figure}[htb]
\includegraphics[angle=-90,width=1\textwidth]{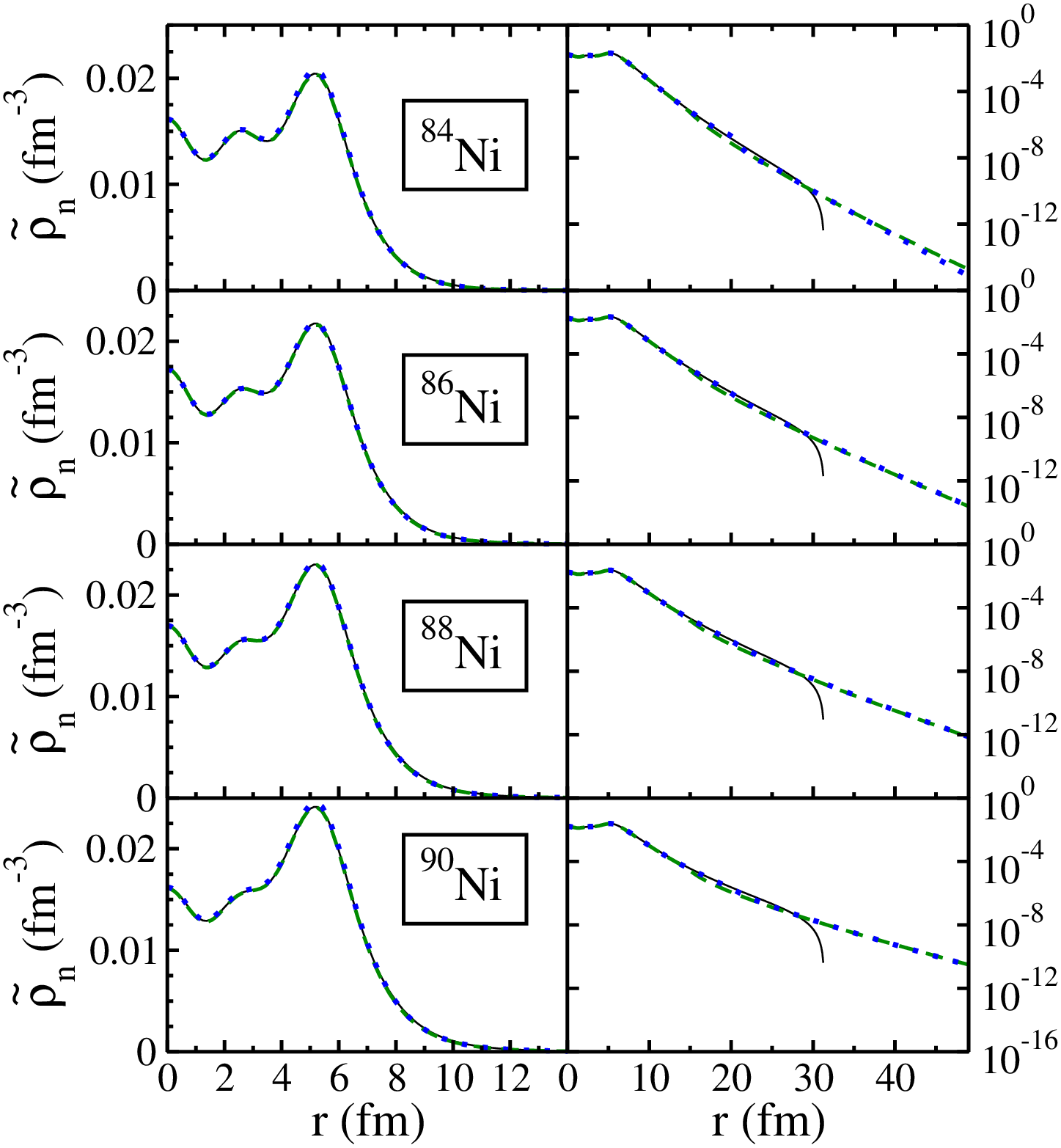}
\protect\caption{\label{pair_dens_Ni}  (color online) 
Same as Fig.~\ref{dens_Ni} but for neutron pairing densities.}
\end{figure}

The result of calculation for normal and pairing densities are presented 
in Figs.~\ref{dens_Ni} and \ref{pair_dens_Ni}.
It is interesting to compare the densities obtained by solving 
the Gamow-HFB equations in the coordinate or the GHF configurational space 
to those calculated by the standard coordinate space framework 
where the continuum is discretized with box boundary conditions. 
They are denoted GHFB/Coord.,~GHFB/Config.~and HFB/Box., respectively.
All results coincide in both normal and logarithmic scales for $r <$ 30 fm. 
It was also checked that spurious imaginary parts of densities, 
caused by the discretization of the continuum of complex energy,
were negligible, of the order of $10^{-6}$ [fm$^{-3}$] for GHFB/Coord.~and 
$10^{-12}$ [fm$^{-3}$] for GHFB/Config.,~ as the largest error values.
In Table~\ref{neutron_resonant_energies}, 
the bound and resonant single-particle states obtained by the GHF calculation 
are compared with the corresponding quasiparticle states 
calculated by the GHFB/Coord.~method. 
It is obvious that bound HF states can give rise to 
unbound quasiparticle states carrying a sizable width 
when pairing correlations are switched on.

Physical observables associated with the HFB ground states are 
provided in Tables~\ref{Ni84_Ni86_res_table} and \ref{Ni88_Ni90_res_table}.
On the one hand, differences occur for neutron pairing energies, 
which are most sensitive to continuum effects \cite{MarioUmar}. 
While those of GHFB/Coord.~compared to HFB/Box remain 
of the order of 500 keV, the difference
between GHFB/Config.~and HFB/Box pairing energies can be $\sim$1.5 MeV. 
On the other hand, the r.m.s.~radii and total energies are basically the same,
with a discrepancy of at most $\sim$300 keV for the latter. 
These results indicate that the GHFB/Coord., GHFB/Config.~and 
HFB/Box treatments are all reliable methods to solve the HFB equations 
taking the continuum effects into account. 
As resonant states are explicitly treated in the Gamow HFB approach, 
this implies that the resonant effects can be well accounted for 
also by means of the HFB/Box method. 
This point is not necessarily widely accepted \cite{giai}.
Even though the good agreements 
among the results of the GHFB/Coord., GHFB/Config.~and HFB/Box calculations might be surprising, we see no reason
to suspect that this is an exceptional case valid only for the Ni isotopes considered here.
It will be interesting to examine this point further.

\section{Perspectives for describing decaying nuclei and 
beyond-mean field approaches}
The GHFB/Coord.~method directly provides quasiparticle wave functions 
without using any intermediate basis states. 
Hence, it may be used also to describe decaying nuclear ground states 
in the HFB approximation. 
In fact, no HFB theory capable of describing decaying HFB ground states exists, 
even though an approximate scheme was proposed in Ref.~\cite{Kyoto_michel}. 
The main difficulty is that it is not possible to construct the HFB
ground state obeying the outgoing wave condition 
if one includes the full set of quasiparticle states of positive energy 
\cite{Kyoto_michel}. 
This arises from the fact that quasiparticles form a degenerate continuum 
of scattering states for $E < |\lambda|$ if $\lambda > 0$, 
whereas they can only generate a discrete set of bound states in this region 
if $\lambda < 0$.
It is impossible to remove quasiparticle states with $E < |\lambda|$  
with the use of the GHFB/Config.~method,
because quasiparticle eigen-energies of the HFB matrix are complex.
In contrast, the direct integration method (GHFB/Coord.) allows us 
to select which quasiparticle states are occupied in the HFB ground state. 
Hence, it may be possible to carefully study properties of decaying 
HFB states at least for the spherical case.

The GHF configurational approach (GHFB/Config.) may be more appropriate 
to study excited states in deformed nuclei by means of the QRPA.
For deformed nuclei, basis expansion approaches may be 
easier compared to the calculation of deformed HFB ground 
states in coordinate space \cite{Obe03}.
For calculating bound HFB ground states, 
we can use the PTG basis, which is more efficient than the GHF basis, 
considering the numerical cost of recalculating the GHF basis states 
inherent to the two-basis method 
(see Sec.~\ref{two_basis_chapter}).
Once a HFB ground state is obtained in this way, one can readily 
calculate the GHF basis wave functions.
The QRPA matrix would then be represented afterward with 
respect to the quasiparticles wave functions expanded in the GHF basis, 
thus allowing the description of unbound QRPA excited states.
%A simple approach utilizing Gamow states with RPA has already 
%been successfully applied within resonant RPA formalism \cite{RRPA}.

\section{Conclusion}
The Berggren completeness relation, 
originally developed in the context of standard Schr\"odinger equation, 
has been extended to quasiparticles in the HFB formalism.
It was shown that, as in the standard single-particle potential problem, 
bound, resonant and scattering quasiparticles are well defined
and form a complete set, by which bound HFB ground states can be constructed. 
Both situations are very similar and can be treated by contour deformation 
of continuous real sets of states, even though physical interpretation of
resonant quasiparticles is different from that of resonant single-particles. 
Numerical applications have been effected with neutron-rich Nickel isotopes 
close to the drip line, for which continuum coupling is important.
It was shown that the Gamow-HFB approach, both in coordinate and 
configurational space representations, properly describe densities and 
physical observables. 
Thus, it provides us with an efficient tool to study ground states of 
medium and heavy nuclei close to the drip line. 
With these approaches, QRPA calculation fully taking into account 
continuum coupling may be efficiently carried out.

\section*{Acknowledgments}
The authors acknowledge the Japan Society for the Promotion of Science 
for awarding the Invitation Fellowship for Research in Japan (Long-term) 
to M.~S.~and the JSPS Postdoctoral Fellowship for Foreign Researchers 
to N.~M., which make our collaboration in Kyoto University possible.
This work was supported by the JSPS Core-to-Core Program ``International
Research Network for Exotic Femto Systems,"
and carried out as a part of the U.S.~Department of Energy under 
Contract Nos.~DE-FG02-96ER40963 (University of Tennessee),
DE-AC05-00OR22725 with UT-Battelle, LLC (Oak Ridge National Laboratory), 
and DE-FG05-87ER40361 (Joint Institute for Heavy Ion Research), 
the UNEDF SciDAC Collaboration supported by the U.S.~Department of Energy 
under grant No.~DE-FC02-07ER41457.

\newpage

\begin{table}[ht]
\centering
\caption{Bound and resonant neutron energies and widths for $^{90}$Ni, 
calculated in the GHF approximation and in the GHFB/Coord.~formalism. 
Single-particle energies ($e_i$) and quasiparticle energies ($E_i$) are 
given in MeV and widths ($\Gamma$) in keV.
Note that the GHF $2s_{1/2}$ state dissolves into continuum 
quasiparticle states in the Gamow-HFB description.}

\label{neutron_resonant_energies}
\vspace{.2cm}
\tabcolsep=.2cm
\begin{ruledtabular}
\begin{tabular}{cccccc}
\noalign{\smallskip}
   &   \multicolumn{2}{c}{\text{GHF}}           &   \multicolumn{2}{c}{\text{GHFB/Coord.}}     \\
\noalign{\smallskip}
\noalign{\smallskip}
states     &   $e$ &  $\Gamma$ & $E$ & $\Gamma$ \\
\noalign{\smallskip}\hline \\
$0s_{1/2}$    & -52.618  &   0     & 51.573  & 1.099 $10^{-3}$ \\
$1s_{1/2}$    & -24.630  &   0     & 24.348  & 46.006 \\
$2s_{1/2}$    & -1.196   &   0     & ------  & ------ \\
$0p_{3/2}$    & -41.655  &   0     & 40.796  & 27.282 \\ 
$1p_{3/2}$    & -12.986  &   0     & 12.658  & 490.565 \\
$0p_{1/2}$    & -42.881  &   0     & 38.870  & 27.138 \\
$1p_{1/2}$    & -11.189  &   0     & 10.816  & 404.299 \\
$0d_{5/2}$    & -29.921  &   0     & 29.141  & 0.780   \\
$1d_{5/2}$    & -2.592   &   0     & 3.181   & 194.181 \\
$0d_{3/2}$    & -30.657  &   0     & 25.095  & 22.567 \\ 
$1d_{3/2}$    & -0.349   &   0     & 2.173   & 560.608 \\
$0f_{7/2}$    & -18.177  &   0     & 17.654  & 397.374 \\ 
$0f_{5/2}$    & -11.331  &   0     & 11.065  & 645.638 \\
$0g_{9/2}$    & -6.770   &   0     & 6.570   & 0.807  \\
$0g_{7/2}$    &  1.350   & 6.410   & 3.120   & 63.6131 \\
$0h_{11/2}$   &  3.852   & 52.851  & 5.269   & 131.776 \\
 \noalign{\smallskip}
 \end{tabular}
 \end{ruledtabular}
 \end{table}

\begin{table*}[ht]
 \centering
 \caption{Gamow-HFB observables for $^{84}$Ni and $^{86}$Ni 
 calculated with the GHFB/Coord., GHFB/Config.~and HFB/Box methods.
 The r.m.s.~radii are given in fm and other quantities in MeV. 
 The proton chemical potential $\lambda_p$ is not presented as 
 there is no proton pairing gap.}
\label{Ni84_Ni86_res_table}
 \vspace{.2cm}
 \tabcolsep=.2cm
 \begin{ruledtabular}
 \begin{tabular}{lrrrrrrrrrrrrrrr}
\noalign{\smallskip}
&  \multicolumn{3}{c}{$^{84}$Ni}  &
&  \multicolumn{3}{c}{$^{86}$Ni}  \\
\noalign{\smallskip} &
\multicolumn{1}{c}{\text{HFB/Box}}           &   
\multicolumn{1}{c}{\text{GHFB/Coord.}} & 
\multicolumn{1}{c}{\text{GHFB/Config.}} & \smallskip &
\multicolumn{1}{c}{\text{HFB/Box}}           &   
\multicolumn{1}{c}{\text{GHFB/Coord.}} & 
\multicolumn{1}{c}{\text{GHFB/Config.}} & \smallskip &
\\
\hline \\
$\lambda_n$      &   -1.453  & -1.430   & -1.440   &&   -1.037 & -1.027   & -1.029     \\
$r_n$            &    4.451  &  4.450   &  4.450   &&    4.528 &  4.526   &  4.526     \\
$r_p$            &    3.980  &  3.982   &  3.982   &&    4.001 &  4.001   &  4.001     \\
$\Delta_n$       &    1.481  &  1.535   &  1.564   &&    1.667 &  1.658   &  1.669     \\
$E^{pair}_n$     &  -30.70   & -30.72   & -31.85   &&  -36.52  & -35.85   & -36.39     \\
$T_n$            & 1084.53   & 1086.05  & 1086.46  && 1118.65  & 1118.68  & 1118.78    \\
$T_p$            &   430.47  & 430.23   & 430.17   &&  425.99  & 426.01   & 426.00     \\
$E^{so}_n$       &  -63.379  & -63.164  & -63.01   &&  -61.679 & -61.712  & -61.631    \\
$E^{Coul}_{dir}$ &  132.94   & 132.89   & 132.88   &&  132.26  & 132.25   & 132.25     \\
$E^{Coul}_{exc}$ &  -10.138  & -10.135  & -10.135  &&  -10.084 & -10.085  & -10.085    \\
$E_{tot}$        & -654.89   & -654.89  & -655.05  && -656.933 & -656.836 & -656.971   \\
 \noalign{\smallskip}
 \end{tabular}
 \end{ruledtabular}
 \end{table*}

\begin{table*}[ht]
 \centering
 \caption{Same as in Table~\ref{Ni84_Ni86_res_table} 
 but for $^{88}$Ni and $^{90}$Ni.}
\label{Ni88_Ni90_res_table}
 \vspace{.2cm}
 \tabcolsep=.2cm
 \begin{ruledtabular}
 \begin{tabular}{lrrrrrrrrrrrrrrr}
\noalign{\smallskip}
&  \multicolumn{3}{c}{$^{88}$Ni}  &
&  \multicolumn{3}{c}{$^{90}$Ni}  \\
\noalign{\smallskip} &
\multicolumn{1}{c}{\text{HFB/Box}}           &   
\multicolumn{1}{c}{\text{GHFB/Coord.}} & 
\multicolumn{1}{c}{\text{GHFB/Config.}} & \smallskip &
\multicolumn{1}{c}{\text{HFB/Box}}           &   
\multicolumn{1}{c}{\text{GHFB/Coord.}} & 
\multicolumn{1}{c}{\text{GHFB/Config.}} & \smallskip &
\\
\hline \\
$\lambda_n$      &   -0.671   &  -0.661     & -0.665    &&   -0.342   &  -0.330   &  -0.342     \\
$r_n$            &    4.603   &   4.602     &  4.601    &&    4.677   &   4.674   &   4.675     \\
$r_p$            &    4.021   &   4.022     &  4.022    &&    4.043   &   4.043   &   4.043     \\
$\Delta_n$       &   1.790    &   1.782     &  1.800    &&    1.899   &   1.899   &   1.935     \\
$E^{pair}_n$     &  -41.98    & -41.26      & -42.17   &&   -47.158  &  -46.509  &  -48.449    \\
$T_n$            & 1150.71    &  1150.74    &  1151.02 &&   1182.52  &  1182.91  &    1183.79  \\
$T_p$            &   421.71   &  421.71     &  421.70   &&   417.38   &  417.35   &    417.31   \\
$E^{so}_n$       & -59.558    &  -59.559    & -59.470   &&    -56.898 & -56.887   &   -56.822   \\
$E^{Coul}_{dir}$ &  131.571   &  131.576    &  131.576  &&    130.947 &  130.883  &    130.878  \\
$E^{Coul}_{exc}$ &  -10.033   &  -10.033    & -10.033   &&    -9.980  &  -9.980   &   -9.980    \\
$E_{tot}$        & -658.167   &  -658.082   & -658.272  &&   -658.665 &  -658.635 &   -658.936  \\
\noalign{\smallskip}
\end{tabular}
\end{ruledtabular}
\end{table*}

\begin{thebibliography}{10}

\bibitem{SM_review} 
{E.~Caurier, G.~Martinez-Pinedo, F.~Nowacki, A.~Poves and A.P.~Zuker, Rev. Mod. Phys. \textbf{77}, 427 (2005)}.

\bibitem{Gog75}
{D.~Gogny, Nucl. Phys. A \textbf{237}, 399 (1975)}.

\bibitem{Gir83}
{M.~Girod and B.~Grammaticos, Phys. Rev. C {\bf 27}, 2317 (1983)}.

\bibitem{Egi80}
{J.L.~Egido, H.-J.~Mang, and P.~Ring, Nucl. Phys. A \textbf{334}, 1 (1980)}.

\bibitem{Egi95}
{J.L.~Egido, J.~Lessing, V.~Martin and L.M.~Robledo,
Nucl. Phys. A \textbf{594}, 70 (1995)}.

\bibitem{Dob04}
{J.~Dobaczewski and P.~Olbratowski,
Comput. Phys. Commun. {\bf 158}, 158 (2004)}.

%\bibitem{Riisager}
%{K.~Riisager, Rev. Mod. Phys. \textbf{66}, 1105  (1994)}.

%\bibitem{Kruppa_Kato}
%{A.T.~Kruppa and K.~Kato, Prog. Theor. Phys. \textbf{84}, 1145 (1990)}.

%\bibitem{Csoto}
%{A.~Cs{\'o}t{\'o}, Phys. Rev. C \textbf{49}, 2244 (1994)}.

%\bibitem{Descouvemont}
%{P.~Descouvemont, C.~Daniel and D.~Baye, Phys. Rev. C \textbf{67} 
%044309 (2003)}.

\bibitem{PRL_michel}
{N.~Michel, W.~Nazarewicz, M.~P{\l}oszajczak and K.~Bennaceur, 
Phys. Rev. Lett. \textbf{89} 042502 (2002)}.

\bibitem{PRC1_michel}
{N.~Michel, W.~Nazarewicz, M.~P{\l}oszajczak and J.~Okolowicz, 
Phys. Rev. C \textbf{67} 054311 (2003)}.

\bibitem{PRC2_michel}
{N.~Michel, W.~Nazarewicz and M.~P{\l}oszajczak, 
Phys. Rev. C \textbf{70} 064313 (2004)}.

\bibitem{R_de_la_Madrid1} 
{R.~de la Madrid, A.~Bohm and M.~Gadella, Fortschr. Phys. 
\textbf{50} 185 (2002)}.

\bibitem{R_de_la_Madrid2} 
{R.~de la Madrid, J. Phys. A \textbf{35} 319 (2002); J. Phys. A: Math. Gen. 
\textbf{37} 8129 (2004)}.

\bibitem{sto04}
{M.V.~Stoitsov , J.~Dobaczewski, W.~Nazarewicz, S.~Pittel and D.J.~Dean,
Phys. Rev. C {\bf 68}, 054312 (2003)}.

\bibitem{Bul80}
{A.~Bulgac, Preprint FT-194-1980, Central Institute of Physics, Bucharest,
1980; nucl-th/9907088}.

\bibitem{Dob84}
{J.~Dobaczewski, H.~Flocard, and J.~Treiner,
Nucl. Phys. A \textbf{422}, 103 (1984)}.

\bibitem{Ter03}
{E.~Ter\'an, V.E.~Oberacker and A.S.~Umar,
Phys. Rev. C {\bf 67}, 064314 (2003)}.

\bibitem{Obe03}
{V.E.~Oberacker, A.S.~Umar, E.~Ter{\'a}n
and A.~Blazkiewicz,  Phys. Rev. C {\bf 68}, 064302 (2003)}.

\bibitem{hfbtho}
{M.V.~Stoitsov, J.~Dobaczewski, W.~Nazarewicz and P.~Ring,
Comput. Phys. Commun. \textbf{167}, 43 (2005)}.

\bibitem{PTG_pot} 
{J.N.~Ginocchio, Ann. Phys., \textbf{152} 203 (1984); 
\textbf{159} 467 (1985)}

\bibitem{HFB_PTG_michel} 
{M.~Stoitsov, N.~Michel and K.~Matsuyanagi, 
Phys. Rev. C {\textbf 77}, 054301 (2008)}.

\bibitem{Betan} 
{R. Id~Betan, N.~Sandulescu and T.~Vertse, Nucl. Phys. A 
\textbf{771} 93 (2006)}.

\bibitem{Dussel} {G.G.~Dussel, R. Id~Betan, R.J.~Liotta, T.~Vertse, Nucl. Phys. A 
\textbf{789} 182 (2007)}.

%\bibitem{RRPA} {P.~Curutchet, T.~Vertse, and R.J.~Liotta, 
%Phys. Rev. C \textbf{39}, 1020 - 1031 (1989)}.

\bibitem{Vautherin_Skyrme}
{D.~Vautherin and D.M.~Brink, Phys. Rev. C \textbf{5}, 626 (1972); 
D.~Vautherin, Phys. Rev. C \textbf{7}, 296 (1973)}.

\bibitem{Gal94}
{B.~Gall, P.~Bonche, J.~Dobaczewski, H.~Flocard and P.H.~Heenen,
Z. Phys. A {\bf 348}, 183 (1994)}.

\bibitem{Ter97a}
{J.~Terasaki, H.~Flocard, P.H.~Heenen and P.~Bonche,
Nucl. Phys. A \textbf{621}, 706 (1997)}.

\bibitem{Yam01}
{M.~Yamagami, K.~Matsuyanagi and M.~Matsuo,
Nucl. Phys. A \textbf{693}, 579 (2001)}.

\bibitem{Karim_code} 
{K.~Bennaceur and J.~Dobaczewski, Comp. Phys. Comm. \textbf{168} 96 (2005)}.

\bibitem{Ring_Schuck} 
{P.~Ring and P.~Schuck, 
``The nuclear many-body problem'', (Springer, New York, 1980)}.

\bibitem{Vertse_CS} 
{B.~Gyarmati and T.~Vertse, Nucl. Phys. A, \textbf{160}, 523 (1971)}.

\bibitem{Messiah} 
{A.~Messiah, \textit{Quantum Mechanics}, (Courier Dover, New York, 1999)}.

\bibitem{Dunford_Schwarz} 
{N.~Dunford and J.T.~Schwartz, \textit{Linear operators}, (Wiley Classics Library, New York, 1988)}.

\bibitem{JMP_Michel}
{N. Michel, J. Math. Phys., \textbf{49}, O22109 (2008)}.

\bibitem{Lind} 
{P.~Lind, Phys. Rev. C \textbf{47}, 1903 (1993)}.

\bibitem{Newton_book} 
{R.G.~Newton, \textit{Scattering Theory of Waves and Particles}, 2nd Ed., (Courier Dover, New York, 2002)}.

\bibitem{Bel87}
{S.T.~Belyaev, A.V.~Smirnov, S.V.~Tolokonnikov, and S.A.~Fayans,
Sov. J. Nucl. Phys. \textbf{45}, 783 (1987)}.

\bibitem{Cha98}
{E.~Chabanat, P.~Bonche, P.~Haensel, J.~Meyer and F.~Schaeffer,
Nucl. Phys. A {\bf 635}, 231 (1998)}.

\bibitem{MarioUmar}
{A.~Blazkiewicz, V.E.~Oberacker, A.S.~Umar and M.~Stoitsov,
Phys. Rev. C {\bf 71}, 054321 (2005)}.

\bibitem{giai}
M.~Grasso, N.~Sandulescu, Nguyen Van Giai and R.J.~Liotta,
Phys. Rev. C {\bf 64} 064321 (2001).

\bibitem{Kyoto_michel}
{N.~Michel, W.~Nazarewicz and M.~P{\l}oszajczak,
proceedings of New Developments in Nuclear Self-Consistent 
Mean-Field Theories, YITP, Kyoto, Japan, YITP-W-05-01, B32 (2005)}.


\end{thebibliography}
\end{document}